\documentclass[10pt]{article}
\usepackage{graphicx}
\usepackage{hyperref}

\def\Title#1{\begin{center} {\Large #1 } \end{center}}
\def\Author#1{\begin{center}{ \sc #1} \end{center}}
\def\Address#1{\begin{center}{ \it #1} \end{center}}

\newenvironment{Abstract}{\begin{quotation} \begin{center} 
             \large ABSTRACT \end{center}\bigskip 
      \begin{center}\begin{large}}{\end{large}\end{center} \end{quotation}}

\newenvironment{Presented}{\begin{quotation} \begin{center} 
             PRESENTED AT\end{center}\bigskip 
      \begin{center}\begin{large}}{\end{large}\end{center} \end{quotation}}

\def\beq{\begin{equation}}
\def\eeq#1{\label{#1}\end{equation}}
\def\eeqn{\end{equation}}

\def\beqa{\begin{eqnarray}}
\def\eeqa#1{\label{#1}\end{eqnarray}}
\def\eeqan{\end{eqnarray}}

\let\bar=\overbar

\def\Dslash{\not{\hbox{\kern-4pt $D$}}}
\def\dslash{\not{\hbox{\kern-2pt $\del$}}}

\def\msb{{\bar{\ssstyle M \kern -1pt S}}}

\usepackage{color}

\newcommand{\avrg}[1] {\langle #1\rangle}

\newcommand{\jpsi}{{{\rm J}/\psi}}
\newcommand{\psiprime}{{\psi({\rm 2S})}}
\newcommand{\chic}{{\chi_c}}

\newcommand{\pp}{{\rm pp}}
\newcommand{\ppb}{{{\rm p\textrm{-}Pb}}}
\newcommand{\pbpb}{{{\rm Pb\textrm{-}Pb}}}
\newcommand{\auau}{{{\rm Au\textrm{-}Au}}}
\newcommand{\dau}{{{\rm d\textrm{-}Au}}}

\newcommand{\pt}{{p_{\rm T}}}

\newcommand{\snn}{{\sqrt{s_{\rm NN}}}}
\newcommand{\raa}{{R_{\rm AA}}}
\newcommand{\rppb}{{R_{\rm pPb}}}

\newcommand{\tppb}{{T_{\rm ppb}}}
\newcommand{\npart}{{N_{\rm part}}}

\newcommand{\ylab}{{y_{\rm lab}}}
\newcommand{\ycms}{{y_{\rm cms}}}
\newcommand{\lint}{{L_{\rm int}}}

\def\affiliation{
On behalf of the ALICE Collaboration, \\
CEA/IRFU/SPHN, \\
L'Orme Des Merisiers, 91191 Gif/Yvette CEDEX, FRANCE }

\begin{document}

% large size for the first page
\large
\begin{titlepage}
%\pubblock

%% Change the title, name, abstract
%% Title 
\vfill
\Title{Charmonium Production at Forward Rapidity in $\pp$, $\ppb$ and $\pbpb$ Collisions, with ALICE }
\vfill

%  if you need to add the support use this, fill the \support definition above. 
%   \Author{ FIRSTNAME LASTNAME \support }
\Author{Hugo Pereira Da Costa}
\Address{\affiliation}
\vfill
\begin{Abstract}
This contribution focuses on latest ALICE results on charmonium forward production in proton-proton ($\pp$), proton-lead ($\ppb$) and lead-lead ($\pbpb$) collisions at the TeV scale. In $\ppb$ and $\pbpb$ collisions, measurements are presented in the form of the charmonium nuclear modification factor (the properly normalized ratio of its production cross section in heavy ion collisions to its $\pp$ counterpart) as a function of the charmonium rapidity and transverse momentum. These measurements are compared to available theoretical calculations. Possible interpretations of these results in terms of gluon saturation, initial and final state energy loss, color screening and recombination are also discussed. \end{Abstract}
\vfill

% DO NOT CHANGE 
\begin{Presented}
The Second Annual Conference\\
 on Large Hadron Collider Physics \\
Columbia University, New York, U.S.A \\ 
June 2-7, 2014
\end{Presented}
\vfill
\end{titlepage}
\def\thefootnote{\fnsymbol{footnote}}
\setcounter{footnote}{0}
%

% normal size for the rest
\normalsize 

%% Your paper should be entered below. 

\section{Introduction}

Charmonia are mesons formed out of a charm and anti-charm quark pair. Their production in heavy ion collisions starts very early via the hard scattering of two partons. As such they constitute a prominent tool to probe the properties of the matter formed in such collisions and in particular the plasma of quarks and gluons (QGP), a state of this matter for which quarks and gluons are deconfined and can travel distances much larger than the typical size of the nucleon. Quarkonia have therefore been extensively studied over the past 30 years for instance at the SPS (CERN), at RHIC (BNL) and more recently at the LHC (CERN)~\cite{Brambilla:2010cs}. In absence of a QGP, quarkonia carry information about the partonic structure of the nucleon, be it isolated or inside a nucleus, and the mechanisms by which partons and heavy quarks lose energy inside cold nuclear matter. In presence of a QGP on the other hand, their production was originally predicted to be suppressed with respect to expectations based on production rates in proton$-$proton ($\pp$) collisions, due to a Debye-like color screening mechanism~\cite{Matzui:1986}. Additionally, it has later been conjectured that their production could also be enhanced due to the recombination of uncorrelated heavy quark pairs from the hot medium~\cite{Thews:2000rj,BraunMunzinger:2000px}. Both the suppression and recombination rates depend on the nature, rapidity ($y$) and transverse momentum ($\pt$) of the charmonium at hand, as well as the energy and the centrality of the heavy ion collision. Measurements performed at LHC energies complement the observations collected at lower energies and help disentangle between these competing mechanisms. 

ALICE~\cite{Aamodt:2008zz} has measured the production of two charmonium states, namely $\jpsi$ and $\psiprime$, down to zero $\pt$ at forward rapidity ($2.5<\ylab<4$) in $\pp$, $\ppb$ and $\pbpb$ collisions. Charmonia are measured in the dimuon-decay channel using ALICE's forward muon tracking system (MCH) and muon trigger system (MTR) in the pseudo-rapidity range $-4 < \eta < -2.5$~\cite{Aamodt:2011gj}. Triggering is achieved using ALICE V0 hodoscopes~\cite{Abbas:2013taa} and the two innermost layers of the Inner Tracking System (ITS)~\cite{arXiv:1001.0502} together with the MTR. The ITS is also used to measure the collision vertex, whereas in $\pbpb$ collisions the V0 hodoscopes are used to evaluate the collision centrality. Charmonia yields are evaluated by fitting the $\mu^+\mu^-$ invariant mass distribution, as reconstructed in the MCH.

The $\pp$ data sample presented in this contribution has been collected in 2011 at a center of mass collision energy $\sqrt{s}=7$~TeV and corresponds to an integrated luminosity $\lint = 1.35\pm0.07$~pb$^{-1}$. The $\ppb$ data sample has been collected in 2013 at a center of mass energy per nucleon-nucleon collision $\snn=5.02$~TeV. Two beam configurations have been used, by inverting the direction of the orbits of the two particle species. These two configurations allow to probe charmonium production in center of mass rapidity ranges $2.03 < \ycms < 3.53$ and $-4.46 < \ycms < -2.96$ respectively, where positive rapidities refer to the situation where the proton beam is travelling towards the forward muon detectors. The data samples collected in each configuration correspond to integrated luminosities $\lint= 5.01 \pm 0.17$~nb$^{-1}$ and $\lint = 5.81 \pm 0.18$ nb$^{-1}$, respectively. Finally, the $\pbpb$ data sample has been collected in 2011 at $\snn=2.76$~TeV. It corresponds to an integrated luminosity $\lint = 68.8 \pm 0.9$(stat.)$\pm2.5$(syst. $F_{\rm norm}$)$^{+5.5}_{-4.5}$(syst. $\sigma_{\pbpb}$)~$\mu$b$^{-1}$.

All measurements of charmonium production performed by ALICE at forward rapidity are inclusive. For $\jpsi$, they contain, in addition to the charmonium direct production, contributions from the decay of higher mass excited states, predominantly $\psiprime$ and $\chic$. For both $\jpsi$ and $\psiprime$, they contain as well contributions from non-prompt production, mainly from $b$-mesons decay. 

\section{Results on $\jpsi$ production at forward rapidity}

\begin{figure}[htb]
\centering
\begin{tabular}{cc}
\includegraphics[width=0.45\textwidth]{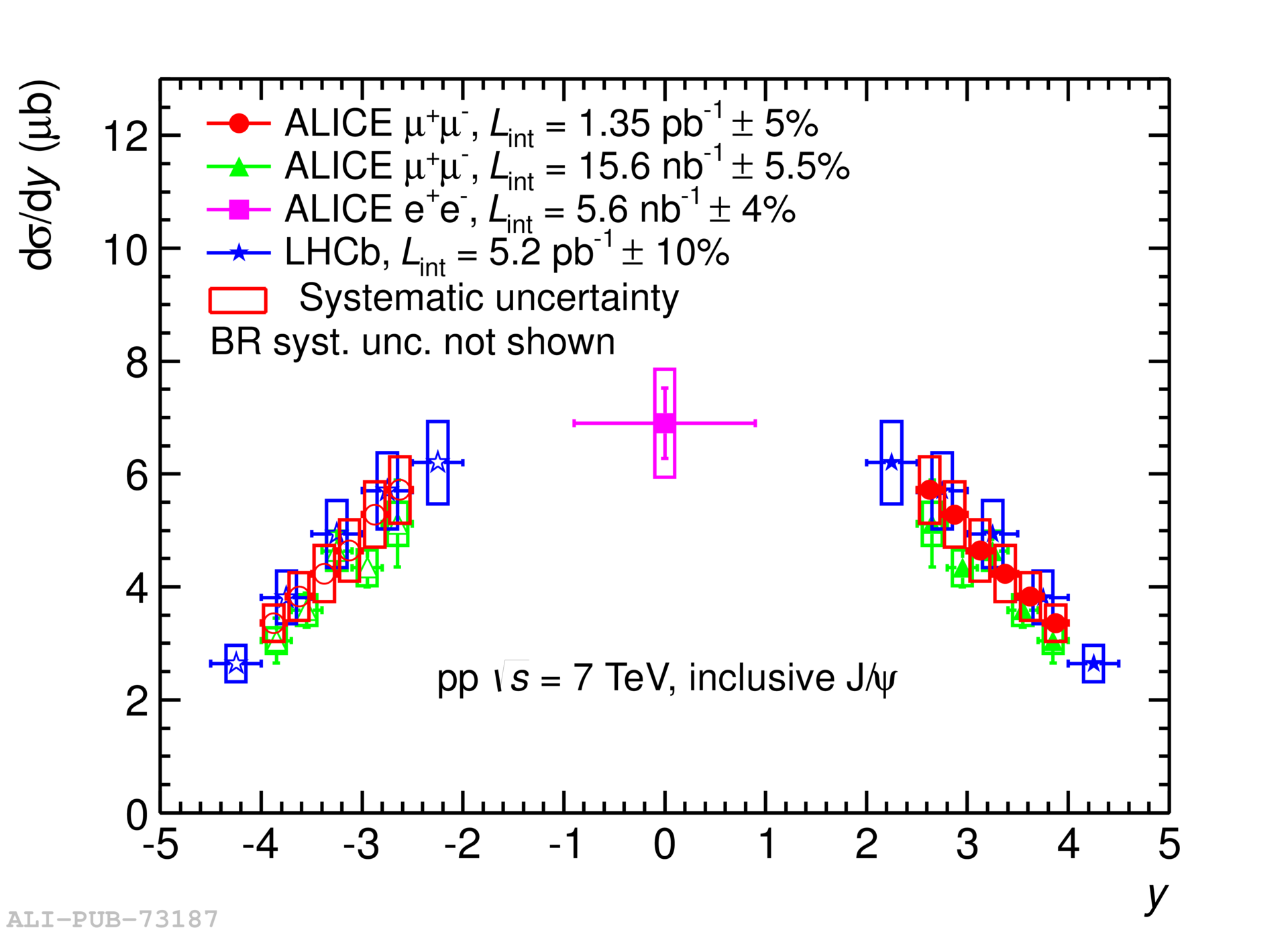}&
\includegraphics[width=0.45\textwidth]{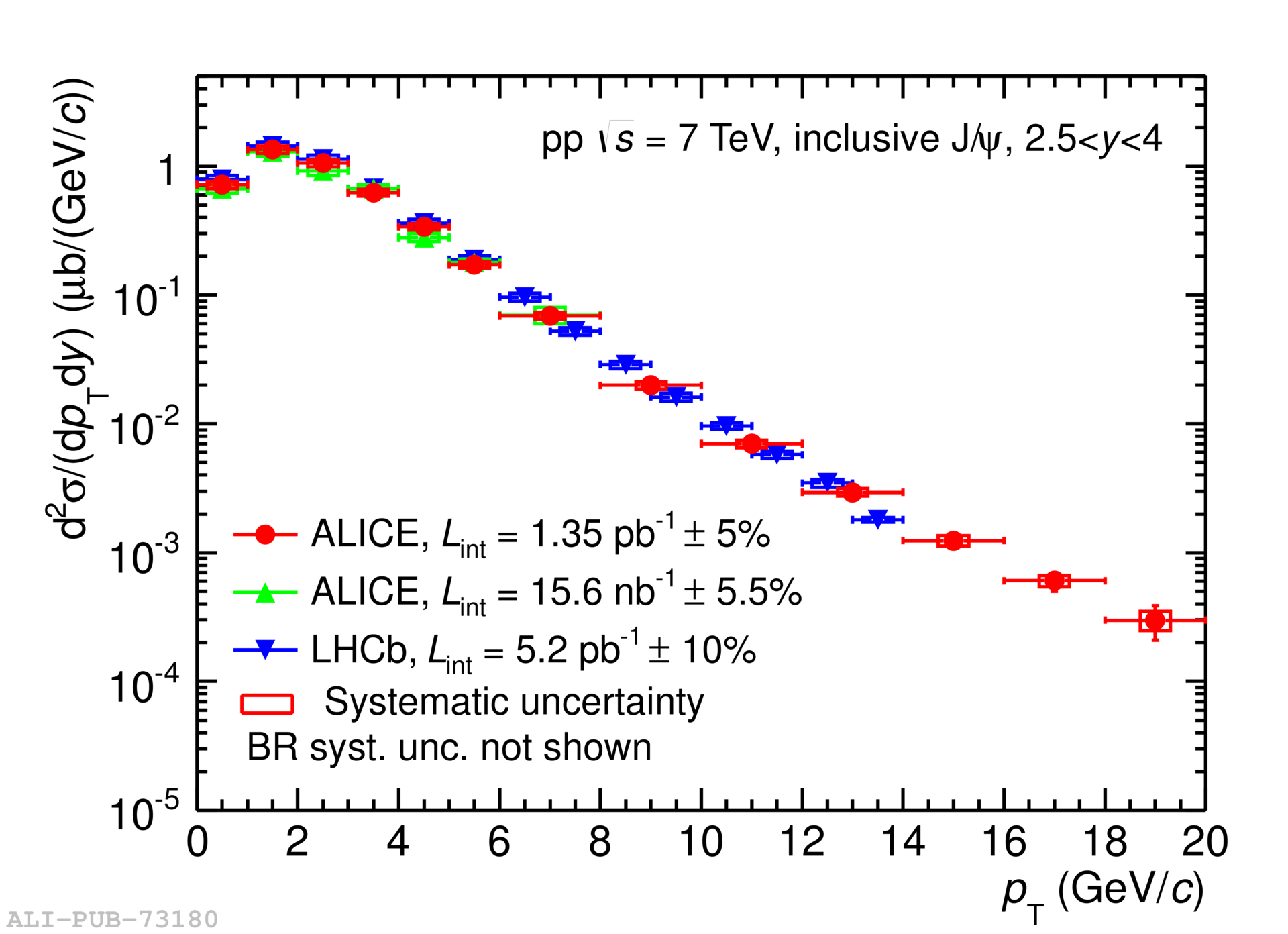}
\end{tabular}
\caption{Inclusive $\jpsi$ production cross section in $\pp$ collisions at $\sqrt{s}=7$~TeV as a function of $\jpsi$ rapidity (left) and $\pt$ (right).}
\label{fig:jpsi_pp}
\end{figure}

Fig.~\ref{fig:jpsi_pp} shows the inclusive $\jpsi$ production cross section measured by ALICE in $\pp$ collisions at an energy $\sqrt{s}=7$~TeV, as a function of the $\jpsi$ rapidity (left panel) and $\pt$ (right panel)~\cite{Abelev:2014qha}. In this figure as well as all others, the vertical bars represent the statistical uncertainties whereas the boxes correspond to the uncorrelated systematic uncertainties. The systematic uncertainty on the integrated luminosity is quoted in the legend. The systematic uncertainty of on the $\jpsi\rightarrow\mu^+\mu^-$ branching ratio ($\sim 1$\%) is not included. The cross sections are compared to LHCb measurements performed in the same conditions~\cite{Aaij:2011jh} as well as earlier ALICE measurements performed using the 2010 data set~\cite{Aamodt:2011gj}, at both mid- ($|y|<1$) and forward rapidities. All measurements agree well within statistical and systematic uncertainties. The interest of these results is two-folded: 1) they provide precise constrain on $\jpsi$ production mechanism, although extra care must be given to the contributions from non-prompt $\jpsi$'s~\cite{Abelev:2012gx} as well as decays from higher mass excited states; 2) they serve as a reference for measuring modifications to this production in proton-nucleus or nucleus-nucleus collisions. 

In order to compare the $\jpsi$ production in $\ppb$ collisions to that in $\pp$ collision, the nuclear modification factor $\rppb$ is used. It is the ratio between the $\jpsi$ invariant yield measured in $\ppb$ collisions and the $\jpsi$ production cross section measured in the same kinematic domain in $\pp$ collisions, further divided by the average nuclear overlap function $\avrg{\tppb}$, estimated using a Glauber model~\cite{ALICE:2012xs}. In absence of nuclear effects, the $\jpsi$ production in $\ppb$ collisions is expected to scale with the number of equivalent nucleon-nucleon collisions and this ratio equals unity.

\begin{figure}[htb]
\centering
\begin{tabular}{cc}
\includegraphics[width=0.45\textwidth]{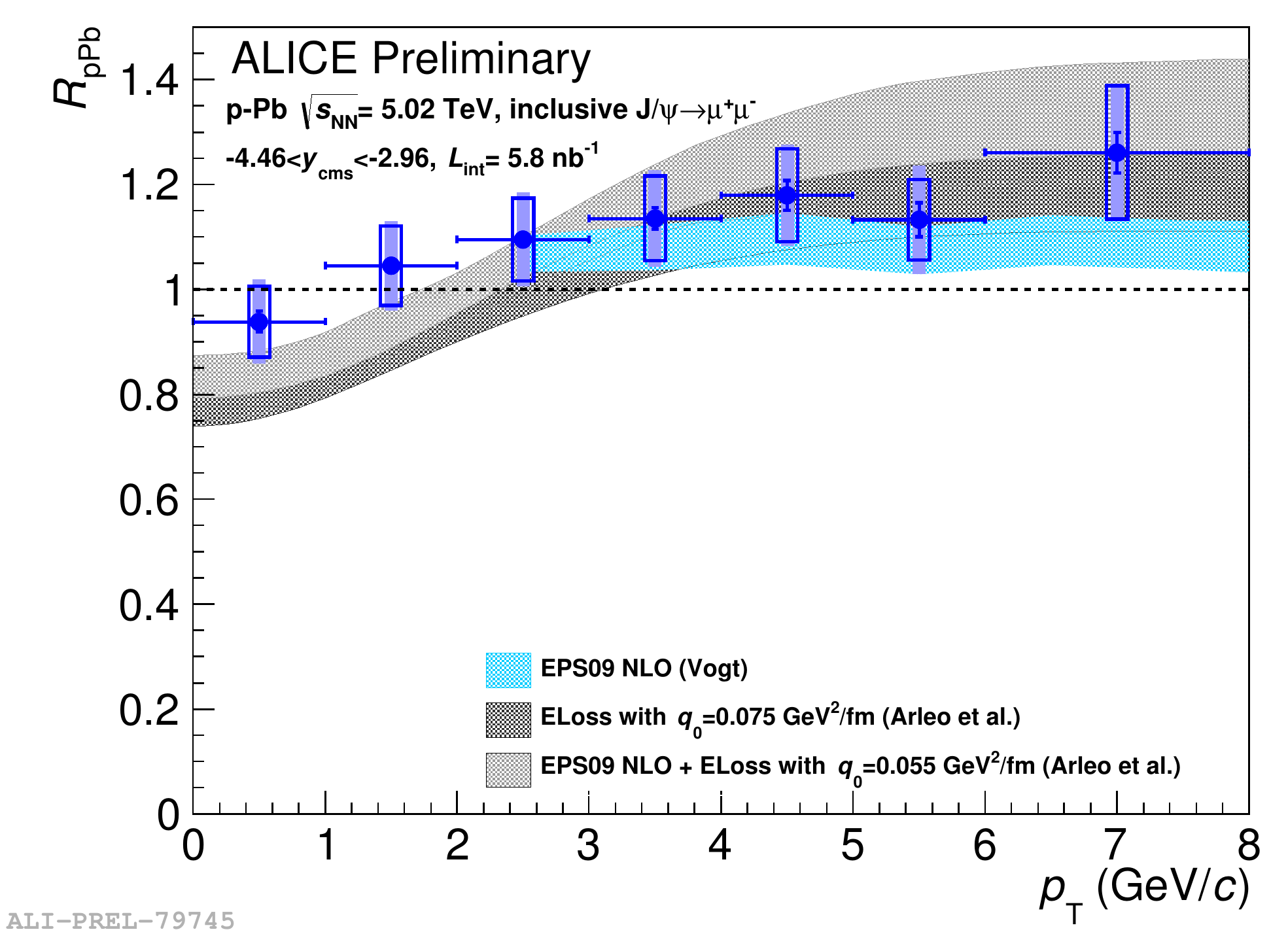}&
\includegraphics[width=0.45\textwidth]{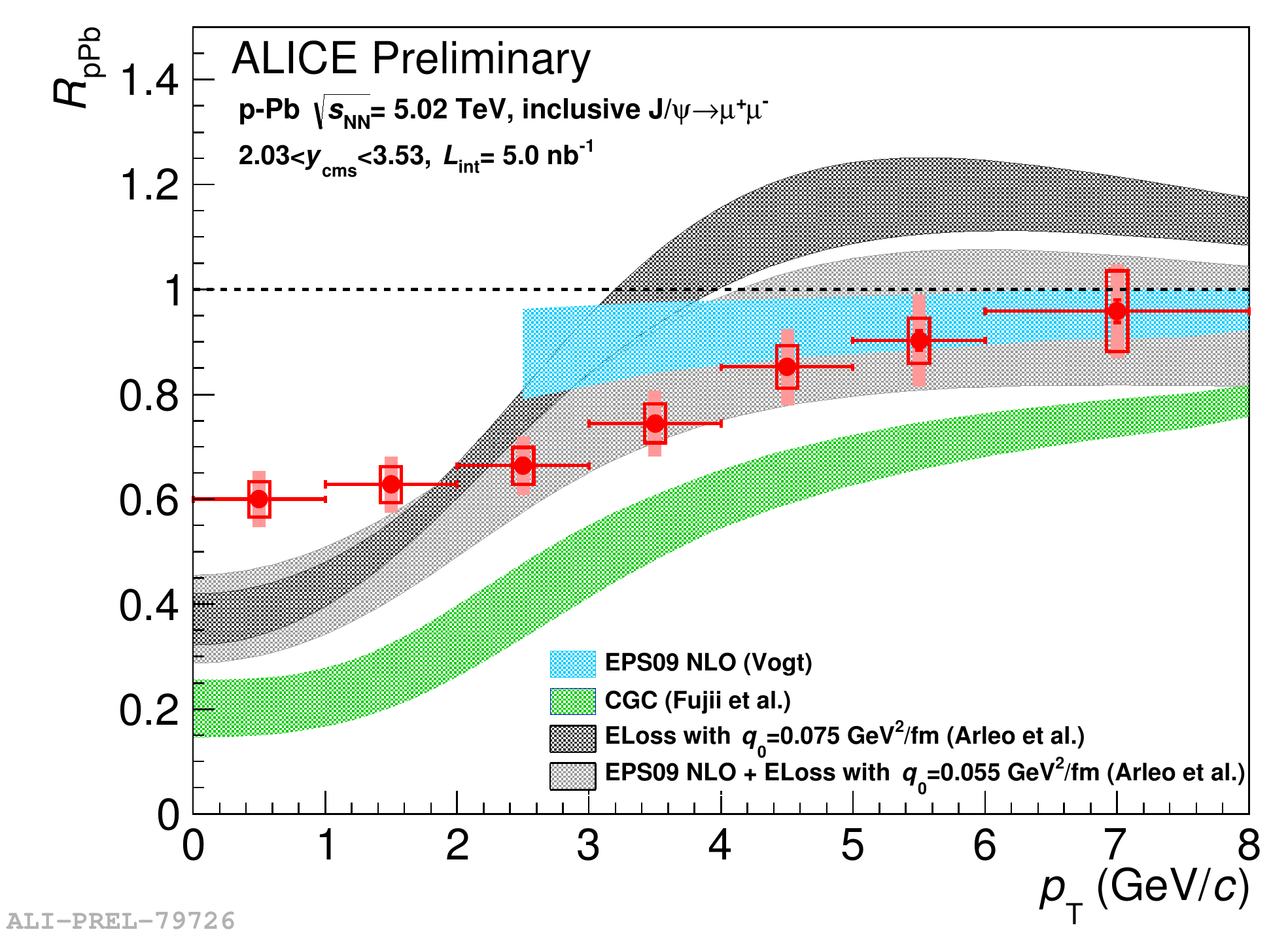}
\end{tabular}
\caption{Inclusive $\jpsi$ nuclear modification factor in $\ppb$ collisions at $\snn=5.02$~TeV as a function of $\jpsi$ $\pt$ at negative (left) and positive (right) rapidity.}
\label{fig:jpsi_ppb}
\end{figure}

Fig.~\ref{fig:jpsi_ppb} shows the $\jpsi$ $\rppb$ measured in $\ppb$ collisions at a center of mass energy per nucleon-nucleon collision $\snn=5.02$~TeV as a function of the $\jpsi$ $\pt$ at negative (left panel) and positive (right panel) rapidity~\cite{Abelev:2013yxa}. Negative rapidity $\jpsi$s originate from gluons that carry a large fraction $x_{\rm Bj}$ of the nucleon longitudinal momentum in the Pb nucleus. Positive rapidity $\jpsi$s on the contrary originate from small $x_{\rm Bj}$ gluons in the Pb nucleus. Such gluons are subject to nuclear shadowing~\cite{Eskola:2009uj}, or saturation, in the framework of the Color Glass Condensate (CGC)~\cite{Kharzeev:2005zr}. The negative rapidity results are compared to three theoretical calculations. The first is a next to leading order (NLO) calculation~\cite{Albacete:2013ei} which uses the Color Evaporation Model (CEM)~\cite{Fritzsch:1977ay} for $\jpsi$ production and the EPS09 parametrization of the modifications of Parton Distribution Functions (PDF) in the nucleus~\cite{Eskola:2009uj}. The other two calculations include a contribution from coherent parton energy loss~\cite{Arleo:2012rs}, either in addition to EPS09 shadowing or as the only nuclear effect. The positive rapidity results are also compared to a calculation performed in the CGC framework combined with the CEM production model~\cite{Fujii:2013gxa}. The first three calculations provide a fair description of the data within statistical and systematic uncertainties and over the $\pt$ range for which they are applicable, whereas the CGC calculation has no predication at negative rapidity and overestimates the suppression observed at positive rapidity over the full $\pt$ range.

\begin{figure}[htb]
\centering
\begin{tabular}{cc}
\includegraphics[width=0.45\textwidth]{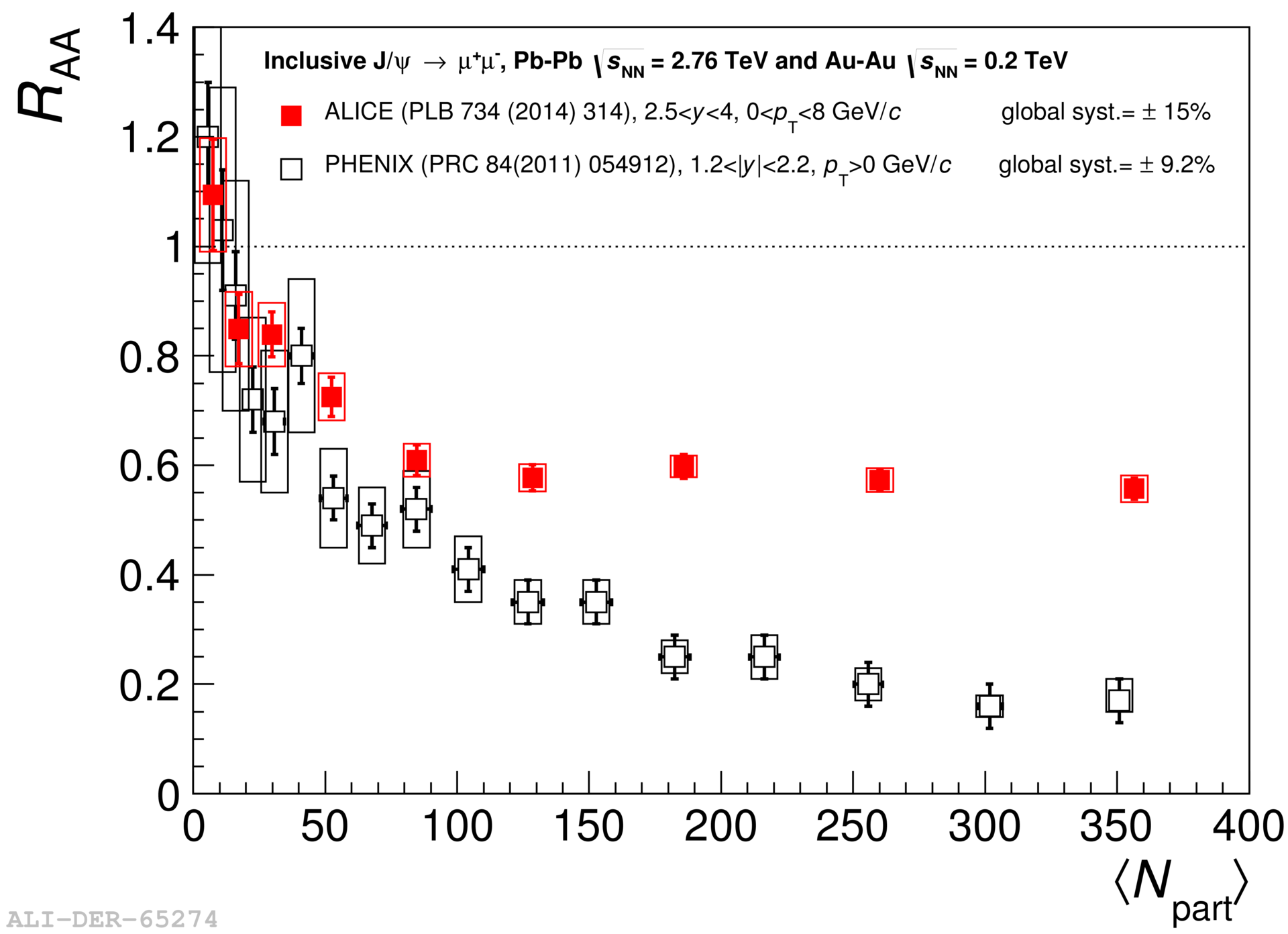}&
\includegraphics[width=0.45\textwidth]{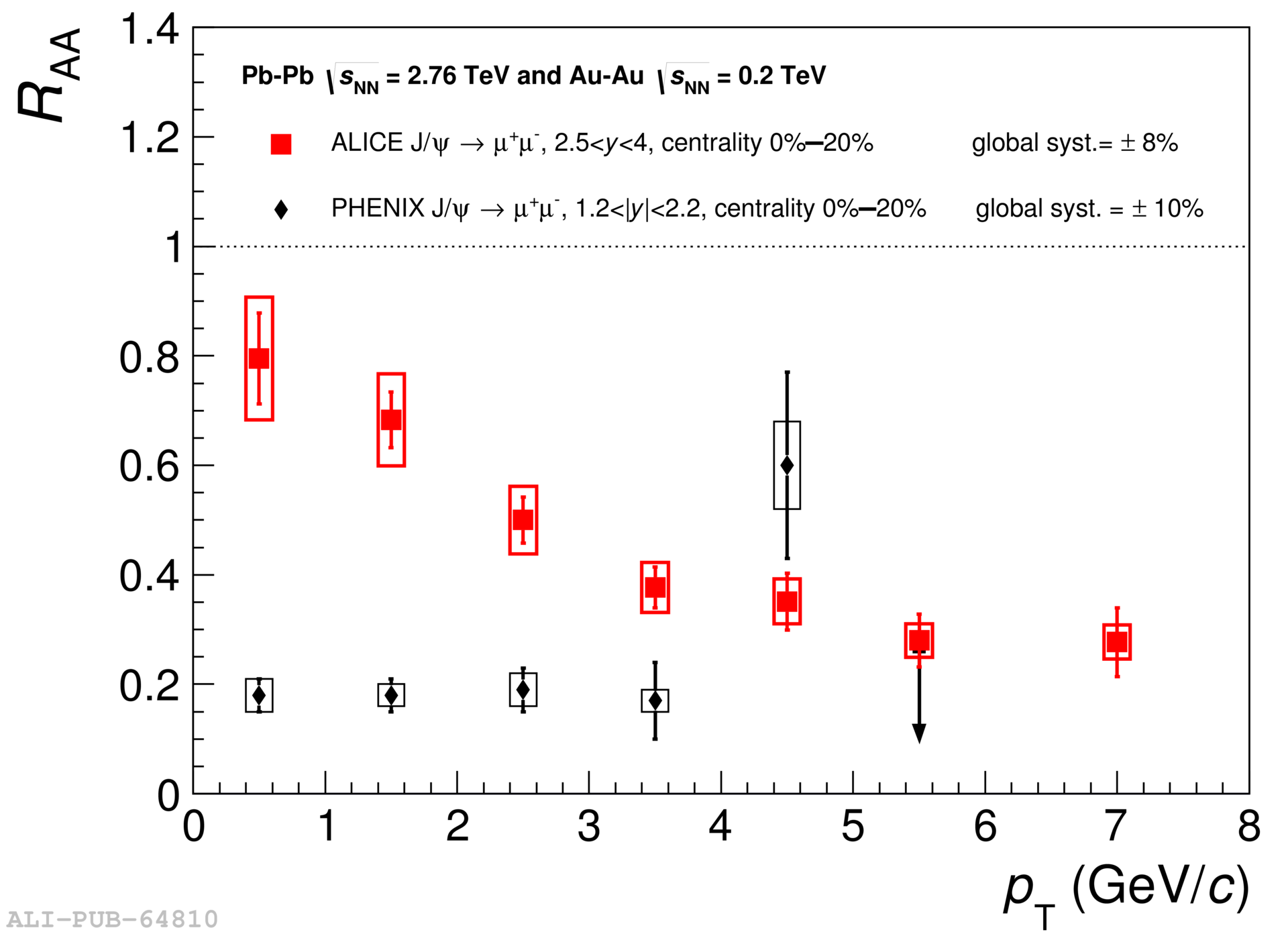}
\end{tabular}
\caption{Inclusive forward rapidity $\jpsi$ nuclear modification factor in $\pbpb$ collisions at $\snn=2.76$~TeV and in $\auau$ collisions at $\snn=0.2$~TeV, left: as a function of the collision centrality; right: as a function of $\jpsi$ $\pt$, for central collisions.}
\label{fig:jpsi_pbpb}
\end{figure}

Fig.~\ref{fig:jpsi_pbpb} shows the $\jpsi$ $\raa$ measured in $\pbpb$ collisions at $\snn=2.76$~TeV as a function of the collision centrality (left panel) and of the $\jpsi$ $\pt$ for central (0-10\%) collisions (right panel)~\cite{Abelev:2013ila}. In the left panel, the collision centrality is characterized by $\npart$ the number of nucleons participating to the nucleus-nucleus collision, estimated using a Glauber model. In both panels the results are compared to measurements performed by PHENIX at RHIC in $\auau$ collisions at $\snn=0.2$~TeV~\cite{Adare:2011yf}. Small values of $\npart$ correspond to peripheral collisions. For such collisions, the nuclear modification factor measured by both experiments is close to unity and little modifications of the $\jpsi$ production is observed with respect to binary scaled $\pp$ collisions. At large values of $\npart$ on the contrary, corresponding to central collisions, a strong suppression of the $\jpsi$ production is observed at both energies. This suppression is however smaller at $\snn=2.76$~TeV than at $\snn=0.2$~TeV. The $\pt$ dependence of the suppression observed for central collisions is also strikingly different between the two collision energies: at high $\pt$, the values of $\raa$ are about the same for both measurements. When $\pt$ decreases however, $\raa$ remains largely unchanged within uncertainties at $\snn=0.2$~TeV, whereas it increases significantly at $\snn=2.76$~TeV. This observation is attributed to the onset of a mechanism that compensates, at high collision energy, the suppression of low $\pt$ $\jpsi$ observed for smaller collision energies. The possibility to produce $\jpsi$  either in the QGP~\cite{Thews:2000rj} or at the phase boundary~\cite{BraunMunzinger:2000px} by the recombination of uncorrelated charm quark pairs out of the medium constitutes one possible scenario for such a mechanism.

\section{Results on $\psiprime$ production at forward rapidity}

\begin{figure}[htb]
\centering
\begin{tabular}{cc}
\includegraphics[width=0.45\textwidth]{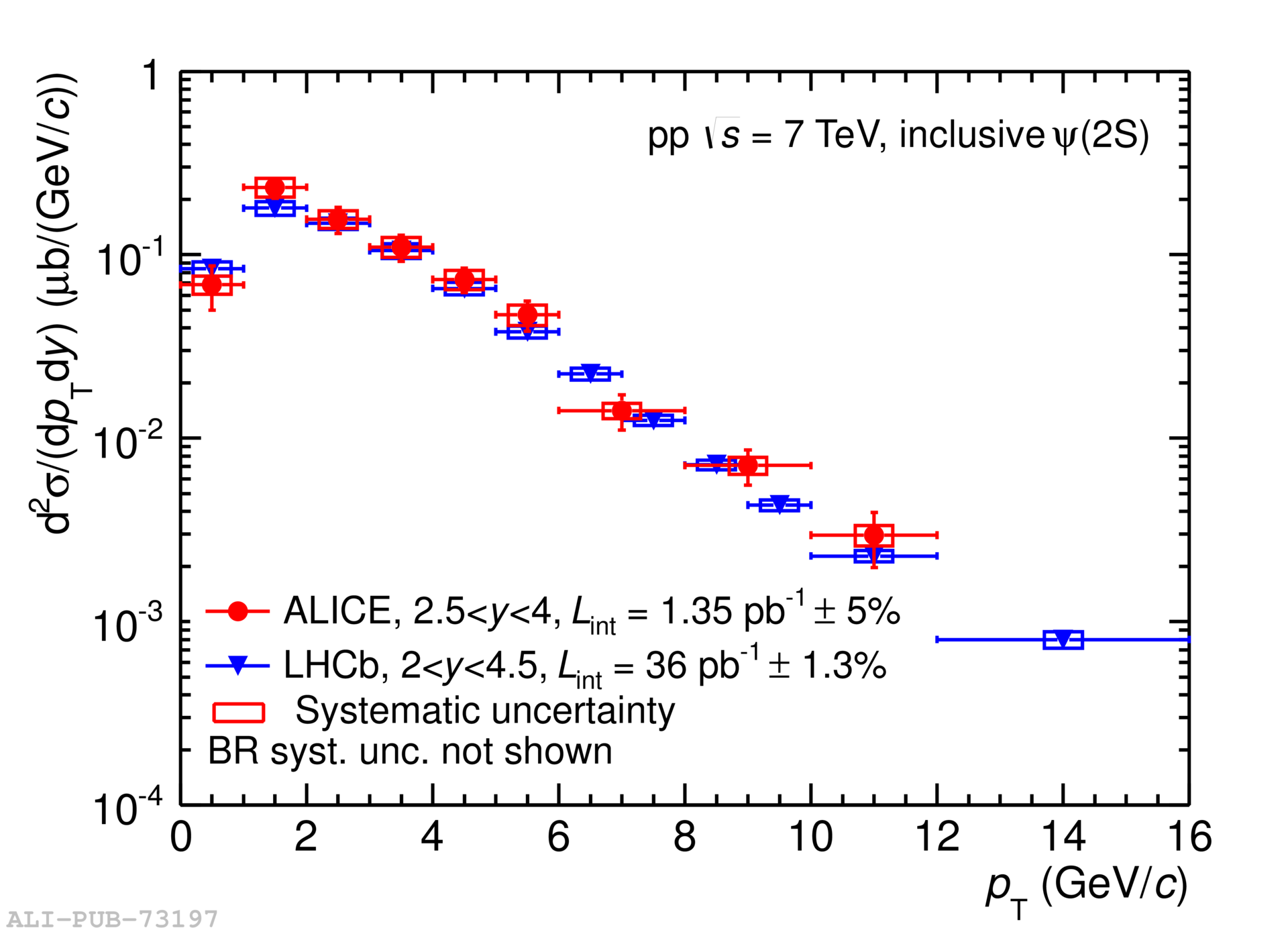}&
\includegraphics[width=0.45\textwidth]{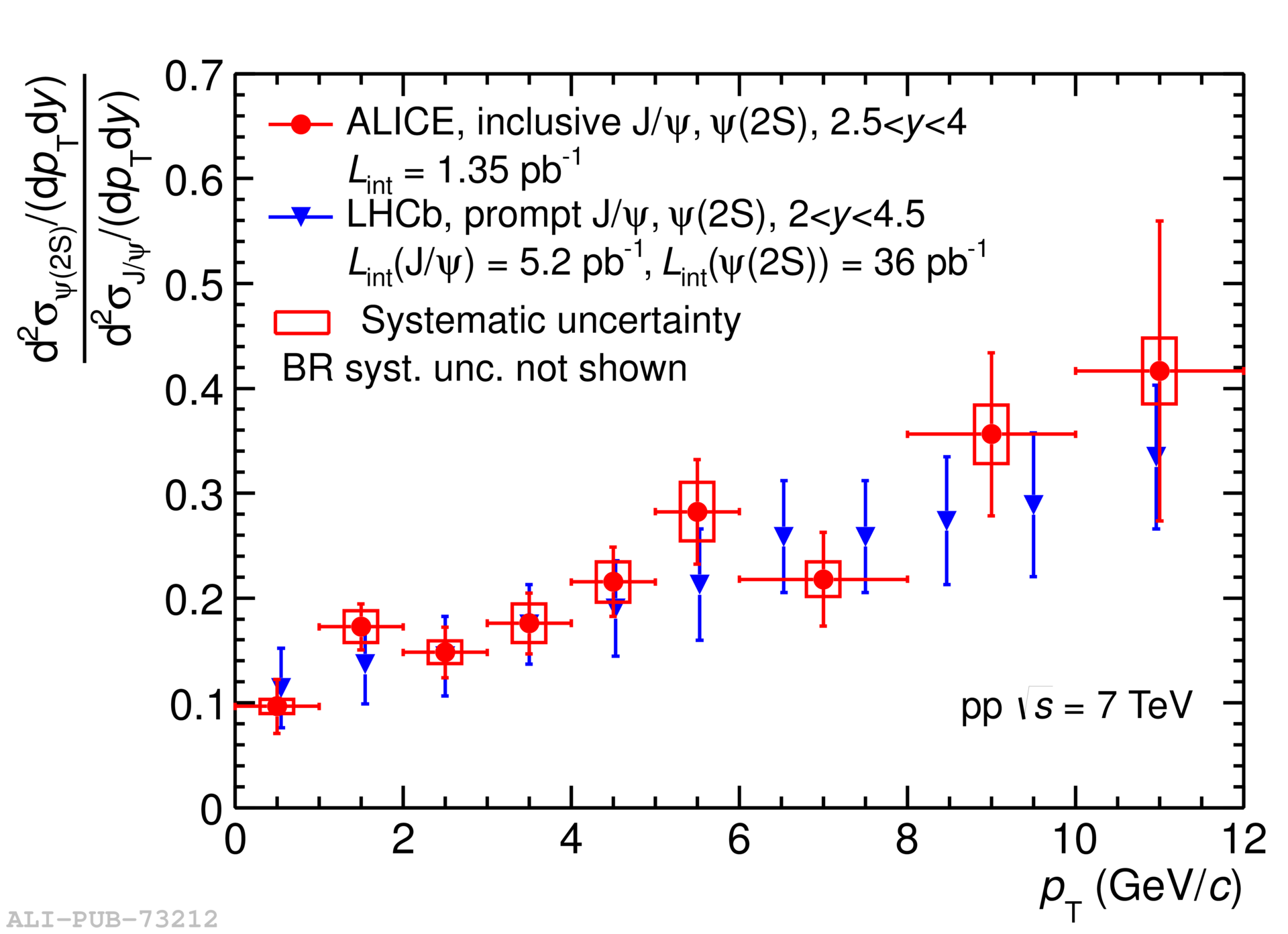}
\end{tabular}
\caption{Left: inclusive $\psiprime$ production cross section in $\pp$ collisions at $\sqrt{s}=7$~TeV as a function of $\pt$; right: inclusive $\psiprime$-to-$\jpsi$ cross section ratio as a function of $\pt$.}
\label{fig:psiprime_pp}
\end{figure}

The inclusive $\psiprime$ forward rapidity production cross section measured by ALICE in $\pp$ collisions at $\sqrt{s}=7$~TeV as a function of the $\psiprime$ $\pt$ is presented in the left panel of Fig.~\ref{fig:psiprime_pp}~\cite{Abelev:2014qha}. It is compared to a measurement performed by the LHCb collaboration in similar conditions although in a slightly different rapidity range~\cite{Aaij:2012ag}. Although both measurements cannot be directly compared, they are in good agreement within uncertainties over the full $\pt$ range. The right panel of Fig.~\ref{fig:psiprime_pp} shows ALICE inclusive $\psiprime$-to-$\jpsi$ cross section ratio as a function of $\pt$~\cite{Abelev:2014qha}, compared to a similar ratio measured by LHCb but for prompt particles only, that is, after the contribution from $b$-meson decays has been taken out. In both cases an increase is observed as a function of $\pt$ which is in contradiction with predictions from the CEM, even after the contribution from higher mass excited states is properly accounted for~\cite{Abelev:2014qha}. These results are used to form the inclusive $\psiprime$ nuclear modification factor in both $\ppb$ and $\pbpb$ collisions as well as the inclusive $\psiprime$-to-$\jpsi$ nuclear modification factor double-ratio.

\begin{figure}[htb]
\centering
\begin{tabular}{cc}
\includegraphics[width=0.45\textwidth]{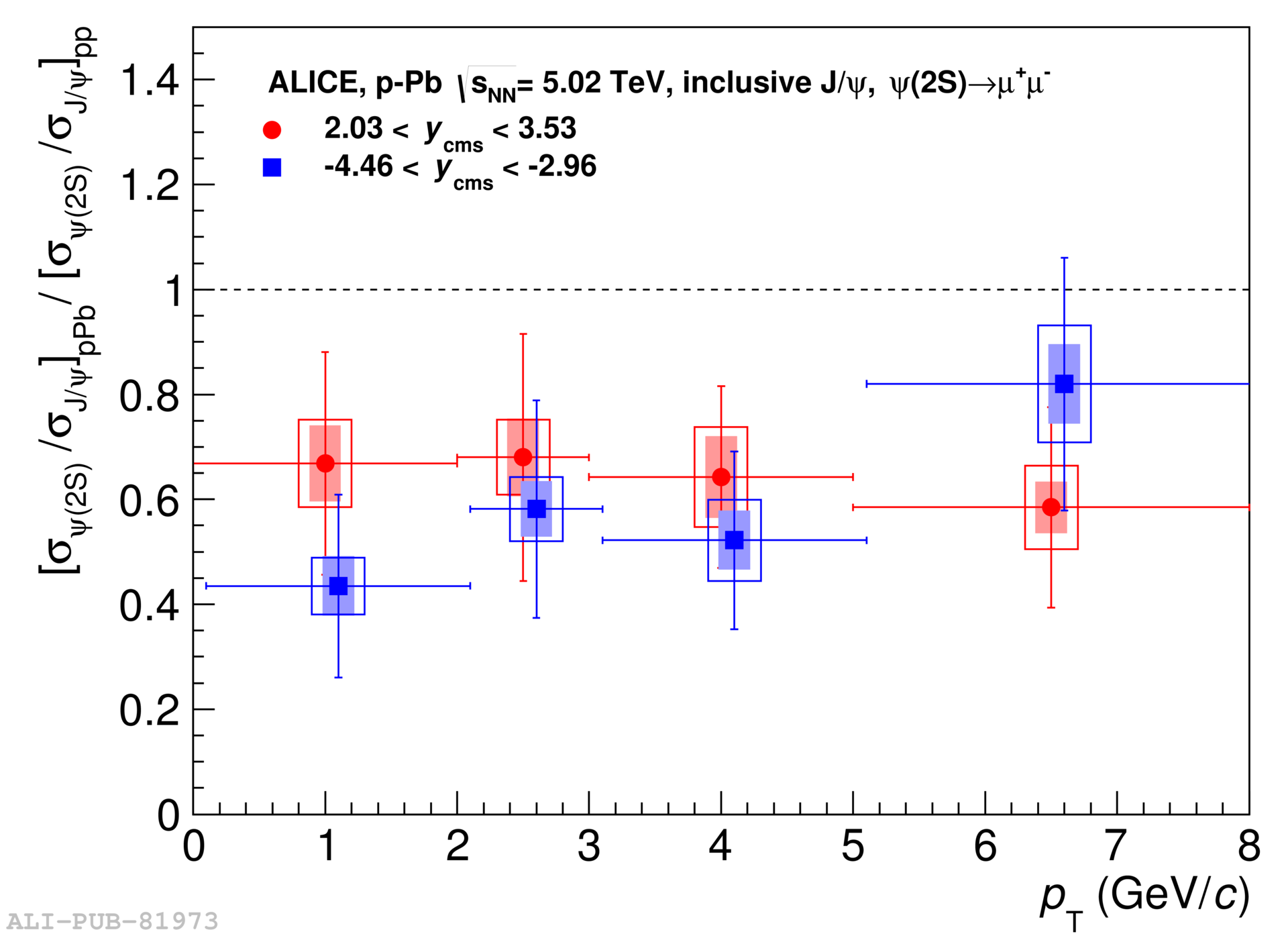}&
\includegraphics[width=0.45\textwidth]{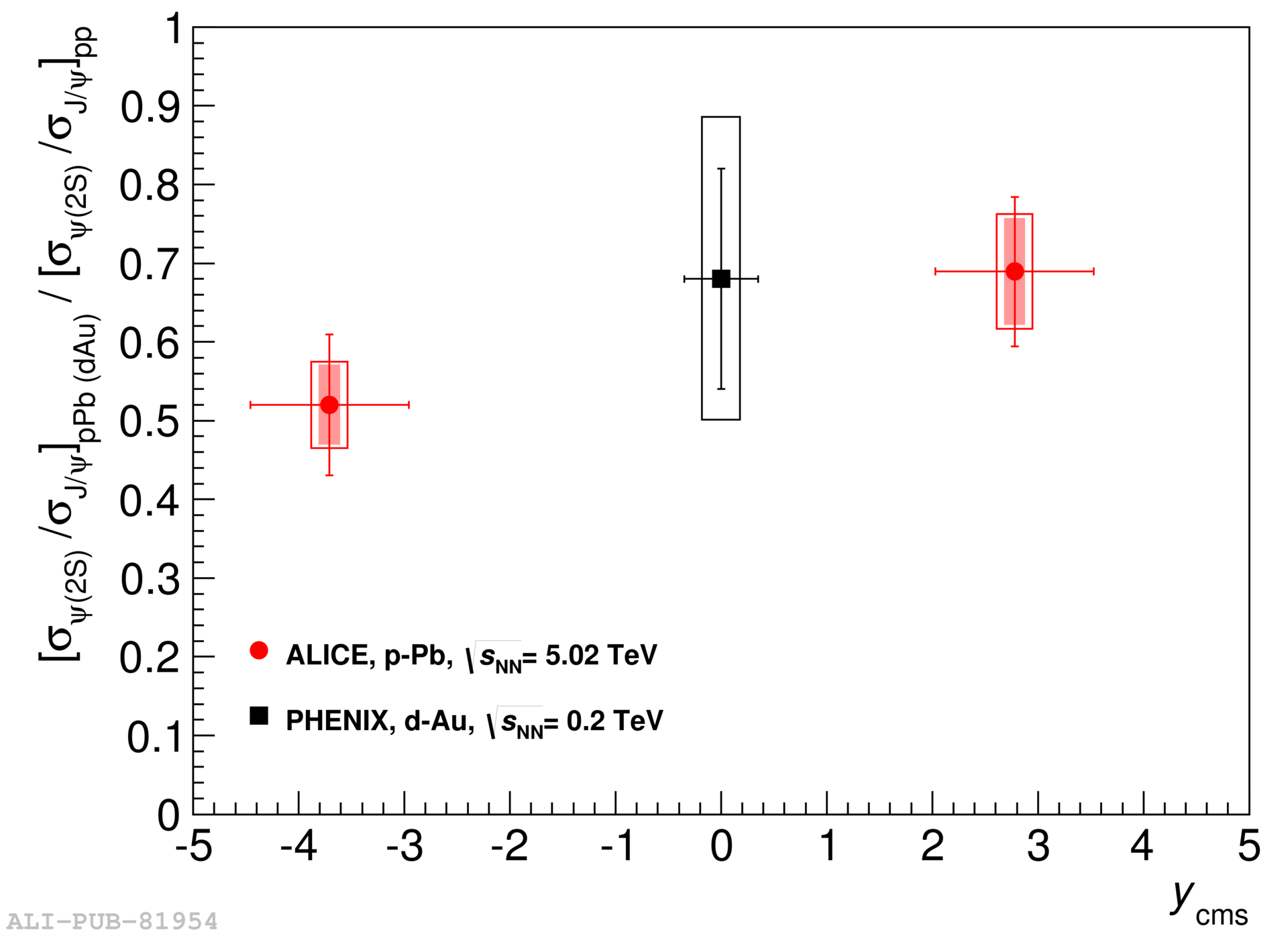}
\end{tabular}
\caption{inclusive $\psiprime$-to-$\jpsi$ nuclear modification factor double-ratio in $\ppb$ collisions at $\snn=5.02$~TeV, left: as a function of $\pt$ at positive (circles) and negative (squares) rapidity; right: as a function of rapidity.}
\label{fig:psiprime_ppb}
\end{figure}

The $\psiprime$-to-$\jpsi$ nuclear modification factor double-ratio measured in $\ppb$ collisions at $\snn=5.02$~TeV as a function of the meson's $\pt$ is shown in the left panel of Fig.~\ref{fig:psiprime_ppb} at both negative- (squares) and positive- (circles) rapidity~\cite{Abelev:2014zpa}. In both rapidity ranges, the ratio is significantly smaller than unity, with no visible dependence upon $\pt$. This indicates that $\psiprime$ mesons are significantly more suppressed than $\jpsi$ in $\ppb$ collisions over the entire rapidity range. In the right panel of Fig.~\ref{fig:psiprime_ppb} these results are compared to a similar measurement performed by PHENIX at mid-rapidity, in $\dau$ collisions and at an energy $\snn=0.2$~TeV~\cite{Adare:2013ezl}. A similar effect is also observed there, although with larger uncertainties. The models presented in Fig.~\ref{fig:jpsi_ppb} predict a similar suppression for $\jpsi$ and $\psiprime$ in $\ppb$ collisions as a function of both $\pt$ and $y$, within a few percent due to slightly different input gluon $x_{\rm Bj}$ distributions. This is a consequence from the assumption that final state effects, which would be able to distinguish between $\jpsi$ and $\psiprime$ mesons, play a minor role at LHC energies. The calculated ratio corresponding to Fig~\ref{fig:psiprime_ppb} would therefore sit around unity, which is in disagreement with the data.
 
\begin{figure}[htb]
\centering
\includegraphics[width=0.45\textwidth]{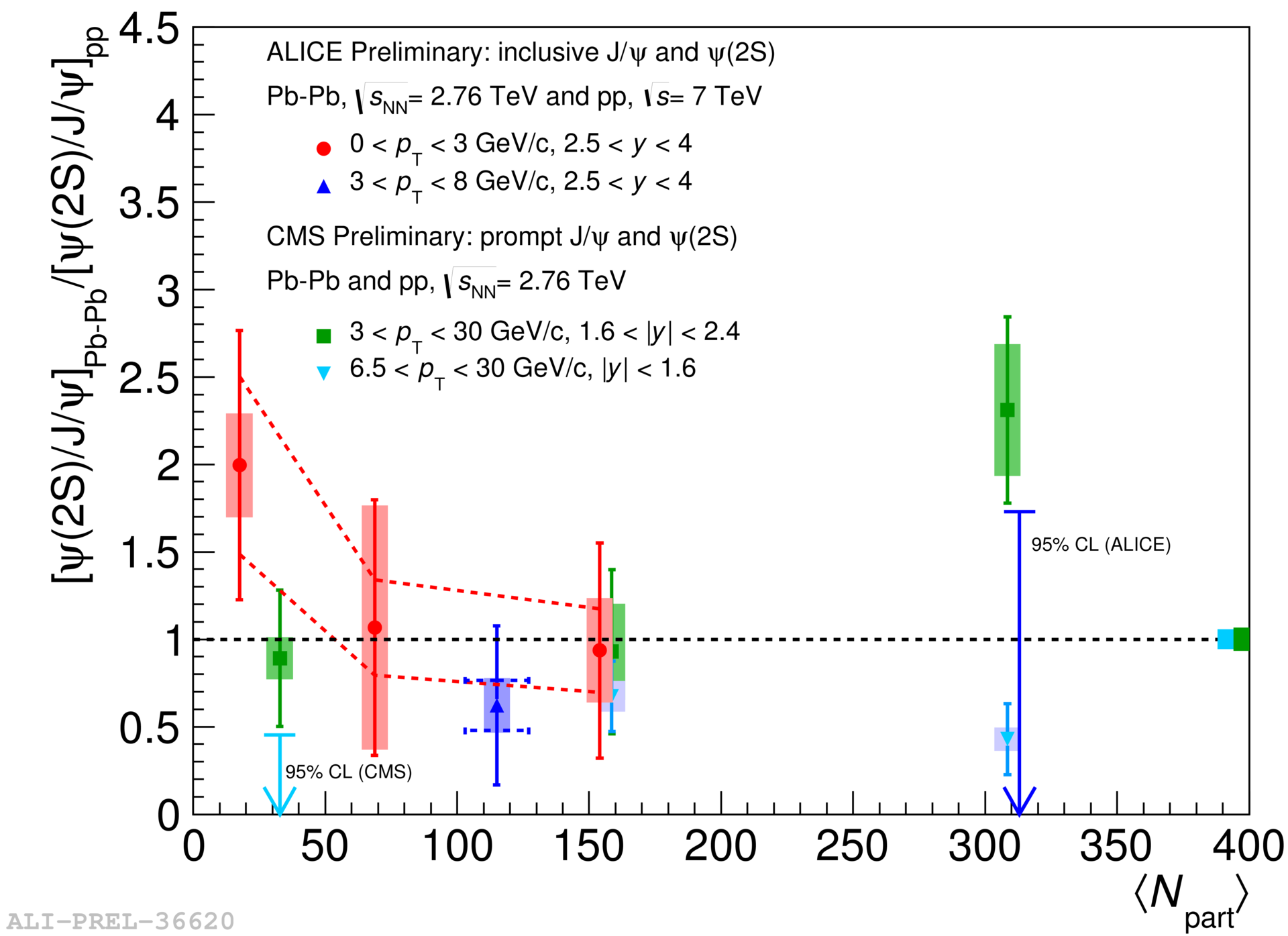}
\caption{inclusive $\psiprime$-to-$\jpsi$ nuclear modification factor double-ratio as a function of collision centrality in $\pbpb$ collisions at $\snn=2.76$~TeV.}
\label{fig:psiprime_pbpb}
\end{figure}

Finally, $\psiprime$ production in $\pbpb$ collision at $\snn=2.76$~TeV has been measured by both ALICE and CMS~\cite{Chatrchyan:2008aa}. The $\psiprime$-to-$\jpsi$ nuclear modification factor double-ratios observed by both experiments are presented in Fig.~\ref{fig:psiprime_pbpb} as a function of the collision centrality ($\npart$), for different $\pt$ and rapidity ranges. For peripheral and semi-central collisions, the double-ratio measured by both experiments is compatible with unity. For most central collisions however, CMS observes large deviations of this double-ratio with respect to unity, and large differences between low-$\pt$ and high-$\pt$ mesons. Presently, too large statistical and systematic uncertainties prevent ALICE measurements to confirm and complement these results at either mid- or forward-rapidity and more data are needed, as will be delivered by the LHC during Run2, starting in 2015.

\section{Conclusion}
In summary, ALICE has measured the inclusive production of $\jpsi$ and $\psiprime$ mesons at forward rapidity in $\pp$, $\ppb$ and $\pbpb$ collisions at center-of-mass energies ranging from $2.76$ to $7$~TeV. In $\pp$ collisions, these measurements put more constrains on charmonia hadro-production mechanism. In $\ppb$ collision, a suppression of the $\jpsi$ production with respect to $\pp$ is observed in the p-going direction. This suppression is more pronounced for low-$\pt$ $\jpsi$s. It is well reproduced, at least qualitatively, by models that include coherent energy loss processes, modifications of the parton distribution functions, or both. For $\psiprime$ a suppression larger than the one measured for $\jpsi$ is observed over the full rapidity and $\pt$ range. This observation is neither expected nor reproduced by models, for which final state effects, that would be able to distinguish between $\jpsi$ and $\psiprime$ mesons, play a minor role at LHC energies. In central $\pbpb$ collisions, a suppression of the $\jpsi$ production is observed that is more pronounced at high $\pt$ than at low $\pt$, in strong contrast to what is observed in $\ppb$ collisions. The suppression at low $\pt$ is also smaller than that observed at lower collision energies by the RHIC experiment. Both observations are consistent with the onset of an additional mechanism such as the production of charmonia in the QGP or at phase boundary by the recombination of uncorrelated charm quarks, and which would partially compensate the suppression observed at lower energies and high $\pt$. Finally, concerning $\psiprime$ production in $\pbpb$, more statistics are needed for a clear picture of the mechanism at play to emerge.

\end{document}